# Trends in Practical Student Peer-review

Helen C. Purchase[1][0000-0001-6994-4446] and John Hamer[2][0000-0001-7051-518X]

[1] Faculty of Information Technology
Monash University
Clayton, Victoria, Australia

[2] School of Computing Science
University of Glasgow
Glasgow, Scotland, UK

**Abstract.** While much of the literature on student peer-review focusses on the success (or otherwise) of individual activities in specific classes – often implemented as part of scholarly research projects  –  there is little by way of published data giving an overview of the range and variety of such activities as used in practice. As the creators, administrators and maintainers of the Aropä Peer review tool, we have unique access to meta-information about peer-review assessments conducted in classes in institutions across the world, together with the variety of class sizes, subjects, rubric design etc. We reported on some of the key assessment configuration data in a 2018 publication covering a period of eight years; here we provide an update on this data – five years later – with particular comment on trends, academic discipline coverage and the possible effect of online delivery during the COVID-19 pandemic.

## 1    Introduction

The benefits of student-peer review are well known: fostering critical thinking (Bhalerao & Ward (2001)), metacognitive self-awareness (e.g. Topping (2005, 2009), Nicol (2010)), self-reflection (e.g. Mulder et al (2014), Harland et al (2017)), judgement making (e.g. Topping (1998), Nulty (2011)), skills of "giving and accepting criticism" (Mulder et al (2014)), as well as helping in demystifying the marking process. (Mulder & Pearce (2007), Topping (2009)). All these learning benefits are, of course, in addition to the practical benefits of students receiving a wide range of feedback on their work in a timely manner. Here we distinguish between 'peer-review' (students provide feedback on their peers' work) and 'peer-assessment' (students provide information on fellow team-members' contribution to a team-based activity).

Any peer-review activity needs to be carefully designed by the instructor, making decisions such as the length of time students will have to comment on their peers' work, the design of the rubric used by the students, the number of submissions each student will comment on. Most papers reporting on specific individual peer-review activities can only describe the design of a few activities known to the researchers in their local context – they are effectively case-studies. Our Aropä data gathers together the design of many more peer-review activities, designed and implemented in a wide range of global contexts. Our data source thus allows us to provide an overview of the universal practices of peer-review in the form of summary statistics; even a meta-review of the many existing research publications on the topic cannot reveal the patterns of typical practice. It is likely that comparable data sets on practical peer-review are held by commercial Learning Management Systems (as conducted using, for example, Moodle Workshop); this information is not typically disseminated.

For this paper, we analysed the design of 3,253 peer-review activities, conducted using Aropä between January 2009 and June 2023 – 14years worth of peer-review activity. We update the KPI summary statistics published in 2018 (Purchase & Hamer, 2018), and analyse trends in the design of peer-review activities. As a research endeavor that considers naturally occurring data, we have no specific research questions or hypotheses in mind: our aim is to see if there are any dominant assessment design characteristics in practical peer review activities. As in the 2018 paper, we attempt to characterize a typical peer-review assessment, and consider the distribution of peer-review activities across different disciplines.

Of course, our data set cannot capture all peer-review activities conducted, only those facilitated by Aropä. However, as a free, non-commercial tool supported by academics, there are no barriers to its use, and we are confident that the 3,253 assessments are likely to provide an almost-all coverage of typical peer-review assessment design. This is useful because it summarises what is actually happening in classes, thus providing pointers to other instructors when they design a new activity or reflect on one that they have completed. Knowing common norms (and deviations from norms) help describe the design space and highlight the range of options available.

## 2 The Aropä system

The source of our data is the Aropä peer-review system, an online system which has been provided free, worldwide, continually since 2009. It has been voluntarily developed and maintained by two Computing Science academics (the authors of this paper) who are the sole designers, developers and maintainers of the system. We are therefore in a unique position to report on the range and scope of peer-review activity in practice.

Up to 30th June 2023, 3,253 successful peer-review assignments devised by 338 instructors have been supported at 53 institutions across the world, in 40 different subject areas. Over 128,000 students have written reviews on their peers' work using the system.

### 2.1 System functions

Aropä supports the principal peer-review activity: anonymous, randomly allocated peer-reviewing, based on a rubric devised by the instructor, with an interface that allows students to upload their submissions before the submission deadline, write reviews of peers' submissions allocated to them before the review deadline, and then view the feedback given on their own submission by other students. The instructor specifies the dates, the rubric and the author/reviewer allocation method; a pairing between a submission and a reviewer is known as an 'allocation'. Table 1 summarises the features provided in Aropä; these 15 features (designated F1-F15) represent the diversity of assignments that can be conducted in Aropä

| F1 | Submission methods | It is possible for any type (and any number) of arte-facts to be required for submission. *(pdf/word)* |
|----|----|----|
| F2 | Reviewer workload | The instructor can specify the number of reviews that each student needs to complete |
| F3 | Duration | The number of days allowed for students to do their reviews can be specified |
| F4 | Rubric | The marking rubric to be used by reviewers can of any length, and have any combination of radio-button 'closed' responses (choosing one option from a list), and text-box 'open' responses (writing text), in any order. |
| F5 | Anonymity | Authors should not know who their reviewers are, but the flexibility for reviewers to know who the authors are is occasionally useful. *(author identity is not revealed)* |
| F6 | Reviewer Allocation | Student reviewers can be everyone in the class, or only those who uploaded a submission. *(only those who submit are allocated submissions to review)* |
| F7 | Submission categories | If students are working in groups, they can submit their work as a group, with one submission associated with all members in the group. Students may also associate a topic tag (taken from a pre-defined set) with their submission. *(individual, non-tagged submissions)* |
| F8 | Allocation method | If submissions are tagged by topic, the instructor can specify that students only review on the topic relating to their own submission, or only on different topic submissions. Students may be asked to work in a group to write a collaborative review. *(within-tag submissions)* |
| F9 | Adjustments | Allocations can be made manually, or manually adjusted after having been automatically and randomly created by the system. *(automatic)* |
| F10 | Self-review | Students can be asked to review their own work – an additional self-review allocation is added to the initial list of allocations. *(no self-review)* |
| F11 | Feedback to authors | The instructor can specify that students can see both the comments and marks in their reviews, or only the comments – this is useful if an instructor wishes |
| | | authors to focus on qualitative responses rather than numeric ones. *(comments and marks revealed)* |
| F12 | Tutor marking | Instructors and tutors can also review authors' submissions, as part of the review process. *(no tutor marking)* |
| F13 | Mark weighting | Different weightings can be associated with the different options in a set of closed responses (represented as radio buttons in the rubric). *(linear, equally spaced weightings)* |
| F14 | Restricted feedback | As an incentive for students to complete their reviews, the instructor can indicate that reviewers can only see their own feedback if they have done at least one or all their allocated reviews. *(all students can see feedback)* |
| F15 | Second-level activity | Reviews themselves can be marked. A secondary assignment can be created that takes as input the reviews from the primary assignment, and allows students (or tutors) to mark these reviews. A variation in this process enables authors to provide a response to their own reviewers' comments. *(no review marking)* |

**Table 1: Peer review features implemented in Aropä (defaults in parentheses, where applicable).**

## 3      The Aropä data

During the period 1st July 2009 to 30 June 2023, 3,596 successful assignments were conducted – we define 'successful' as assignments where more than half of the review allocations were completed. Our conversations with instructors reveal that in most cases there are seldom explicit summative assessment incentives for students to complete the reviews allocated to them; we discovered that it is actually rare in an assignment for all reviews to have been completed.

Of the 3,596 successful assignments, 339 were based on submissions written by instructors rather than students (using the system for giving students practice in reviewing), and a further 4 were used for review marking, leaves us with 3,253 assignments where peers commented on their fellow-students submissions: 1,043 in the first 8.5yrs, 2,210 in the subsequent 6yrs.

These assignments are contained within a total of 1,691 'courses'. A course is typically one semester or term duration, and is associated with an academic subject and an instructor, and with a list of students enrolled in the class.

### 3.1    Key performance indicators

To demonstrate the extent of use of the system, and its value as a source of peer-review activity data, we present Key Performance Indicators (KPIs). One of our main KPIs is the number of unique students who have written reviews, since Cho & Cho (2011) and Nicol et al (2014) both highlight the dominant learning benefit of writing reviews (rather than simply reading them). The number of instructors who have returned for repeated use is an indication of prior successful use, and, since our intention is to support large classes in particular, class size is important. Table 1 summarizes the KPIs, making pre- and post-July 2017 comparisons.

| | | |
|---|---|---|
| Number of successful peer-review assignments | Since January 2008 | 3,253 |
| | Since July 2017 | 2,210 |
| Number of unique students who have used the system to write at least one peer-review. Note that this metric is calculated per year, and so does not take into account students who may have used Aropä in more than one year. | Since January 2008 | 128,925 |
| | Since January 2018 | 65,318 |
| Number of repeat instructors: that is, instructors who have used the system for more than one peer-review activity | Since January 2008 | 254 |
| | Since July 2017 | 165 |
| Number of higher education institutions with at least one successful assignment | Since January 2008 | 54 |
| | Since July 2017 | 30 |
| Largest class size for one assignment | August 2021, (Economics) | 1170 submissions 3,100 reviews |
| Largest number of reviews written in one assignment | October 2022, (Biochemistry) | 336 submissions 7,603 reviews |
| Total number of completed reviews | Since January 2008 | 602,628 |
| | Since July 2017 | 331,307 |

**Table 2: Aropä KPI usage data (at 30[th] June 2023)**

### 3.2 The typical assignment configuration

In our previous analysis of Aropä data, we reported that the typical Aropä assessment had the following form, based on peer-review data from 1,043 assignments (Purchase & Hamer (2018)).

*"Students submit a single pdf or Word document which represents their own work. After the submission deadline, they are randomly allocated two of their peers' submissions to review anonymously; that is, the authors' names are not revealed. Only those students who have submitted a document are allowed to take part in the reviewing process. The students are given a week to write their reviews, using a rubric that comprises two sets of radio buttons, and one comment box. The values associated with the items in each set of radio buttons are increasing and sequential, and start at 1. After the reviewing deadline, all students who made a submission can see all the reviews that have been written on their work, seeing both the responses given to the radio buttons as well as the comments. The students do not know the identity of their reviewers. Tutors do not review the submissions. This is the only peer-review assignment that the students undertake for this class."*

The inclusion of the data since 30[th] June 2017 only changes this description slightly in that the typical rubric no longer has any radio button ('marking') elements, and comprises simply one comment box – indicative of a move towards providing more qualitative, formative feedback rather than summative assessment.

### 3.3 Trend data

Our trend analysis considers the period between 1[st] January 2090 and 31[st] December 2022, so as cover complete years. The data charts (Fig 3-12) show the trend data for assignments, courses, subjects covered by Aropä assignments, as well as for selected features (as described in Table 1 above).

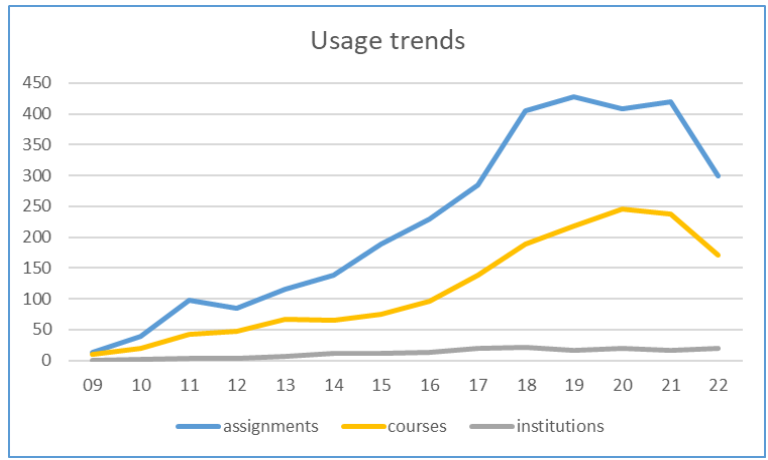

Fig 1: Aropä usage

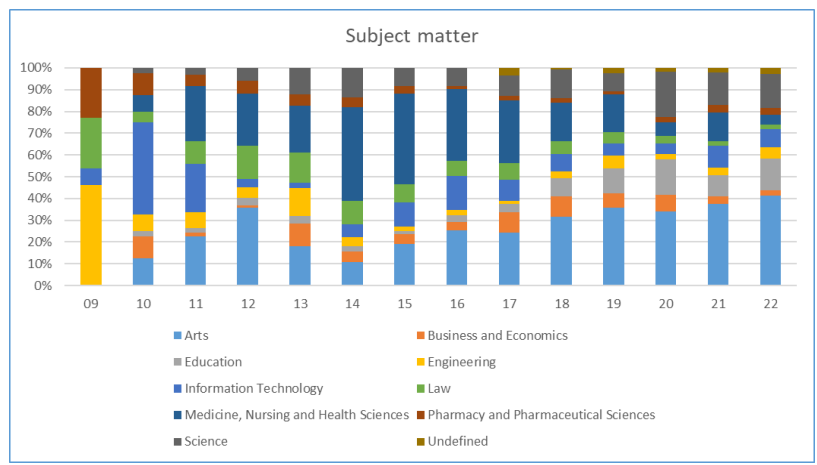

Fig 2: Subject categories of assignments in Aropä (classified according to the Faculties in the author's home institution).

Aropä usage grew steadily since its inception (Fig 1), stayed steady during the COVID-19 pandemic years of 2020 and 2021, but has declined since then. We believe that the reason for this recent decline is the pandemic-related increased use of online tools for educational delivery across the board, resulting in institutions' reassessment of their tools, with a desire for integration of all online functions into a coordinated suite of tools. The consequence might be that the use of an external tool like Aropä is not encouraged.

The relative proportion of the subjects which Aropä assignments cover reveal a continued dominance of Arts and Medicine (Fig 2), steady usage in Law, with in-creases in Education and Science. There was a sudden decrease in Information Technology/Engineering after 2011. Since Aropä was developed by Computing Science/Software Engineering academics, its initial use was predominantly with their colleagues; the outcome of the concerted effort in 2010/11 to make it available in other subjects is evident in this chart.

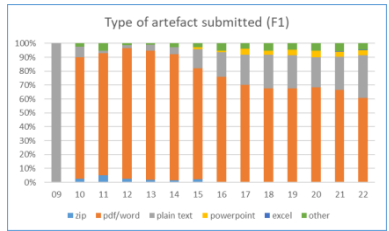

Fig 3: Submission types

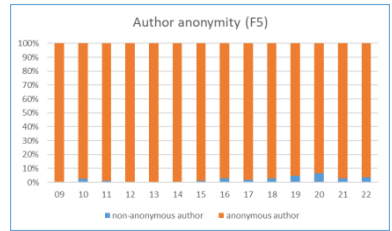

Fig 4: Anonymity

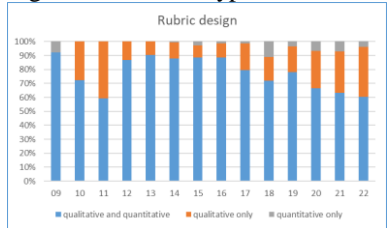

Fig 5: Rubrics

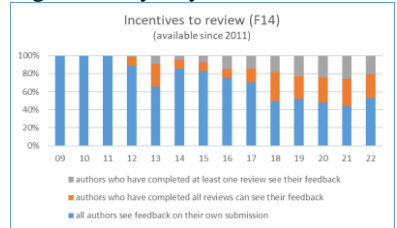

Fig 6: Incentives

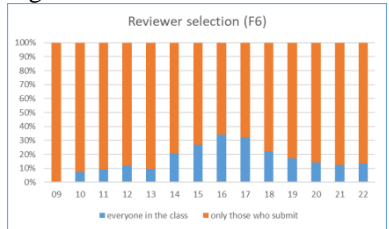

Fig 7: Reviewers

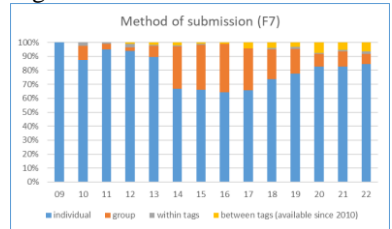

Fig 8: Group/tagged submissions

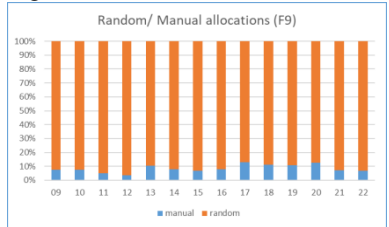

Fig 9: Automatic allocation

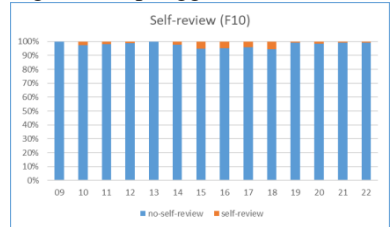

Fig 10: Self-review

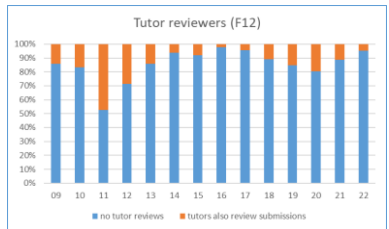

Fig 11: Tutor marking

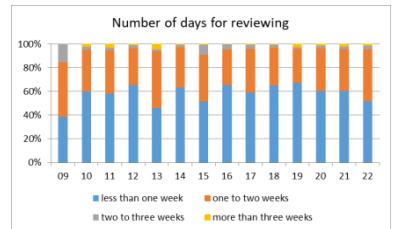

Fig 12: Review period

From the trends in assignment configuration data, we can make the following observations:

- Zip submissions (Fig 3) tend to be associated with Information Technology and Engineering assignments; increasingly instructors are requesting that presentation slides be peer-reviewed as a means of allowing students to provide feedback on student oral presentations.
- Instructors are increasingly allowing reviewers to know the name of the submission author (Fig 4) (although the proportion is still low), suggesting that peer-review us being seen more as a collaborative activity, rather than simply for reviewing or marking.
- The increase in the emphasis in recent years on rubrics that seek only comments (not marks) from reviewers (Fig 5) suggests more focus on the provision of formative feedback; rubrics that only include quantitative marking elements have remained roughly stable in this time.
- In-system incentives (Fig 6) to encourage students to complete the reviews allocated to them have increased over time, peaking during the COVID-19 pandemic years of 2020 and 2021.
- Allowing everyone in the class to review has the risk that students who have not experienced the challenges of completing the assignment may not be able to provide valid reviews (Fig 7); the peak in 2015-2018 relates to extensive repeated activity at one institution where the whole class was asked to comment on the presentations of just a couple of groups of students.
- Instructors frequently comment on the usefulness of the ability for students to submit in groups (Fig 8), although use of this feature has declined in recent years. Using the between-tag option (increased recently) helps mitigate instructors' concerns about potential plagiarism
- Few instructors manually adjust Aropä's automatically randomly created allocations (Fig 9).
- Although self-review is very easy to implement in Aropä (Fig 10), and the benefits of self-review are well known (Nicol et al (2014), few instructors make use of this feature.
- Some early users were concerned about peer-reviewers not giving the 'correct' answer, and so required that tutors also marked the submissions (Fig 11); this is less of the case in recent years.

## 4    Conclusion

This analysis of naturally occurring data from the Aropä peer-review system provides an update to the data presented in our 2018 publication (Purchase & Hamer (2018)), showing where the use of features in the tool has changed over time or remained static. The downward trend in overall use is likely due to post-pandemic efforts by institutions to provide integrated learning tools. We see evidence of instructors moving away from using peer-review as a means of providing summative marks towards its use for formative feedback, and show how online peer-review activities span a wide range of subjects.

**Acknowledgements**. We are grateful for the support of the School of Computing Science at the University of Glasgow and the Department of Computer Science at the University of Auckland for providing hosting services. Ethical approval for this study was given by the Ethics Committee of the College of Science and Engineering, University of Glasgow (ref: 300160176). Aropä can be found at Aropä.gla.ac.uk.